\RequirePackage{snapshot}
\documentclass [11pt, fancyhdr] {article}
\usepackage{float, graphicx, caption, amssymb}
\usepackage[usenames,dvipsnames]{color}
\usepackage{tabulary}
\usepackage [left=2.5cm, top=2.5cm, bottom=2.5cm, right=3cm] {geometry}  
\usepackage{fancyhdr}
\usepackage{xcolor}
\usepackage[scaled]{helvet}
\usepackage{amsthm} 
\usepackage[left]{lineno}
\usepackage[yyyymmdd,hhmmss]{datetime}

\usepackage{blindtext}

\newcounter {note}
\stepcounter{note}

\usepackage{natbib}
\usepackage{amsmath}
\usepackage{amssymb}
\usepackage{bm}
\usepackage{bbm}
\usepackage{tikz}
\usepackage{subcaption}
\usepackage{graphicx}
\usepackage[figuresright]{rotating}
\usepackage{tabularx}
\usepackage{hyperref}
\usepackage[title,titletoc,page]{appendix}

\usepackage{algorithm}
\usepackage{algpseudocode}
\algnewcommand{\Initialize}[1]{%
	\State \textbf{Initialize:} #1
}

\begin{document}

\title{Bayesian non-linear subspace shrinkage using horseshoe priors}
\author {Julia  Christin Duda \textit{duda@statistik.tu-dortmund.de} \\
Department of Statistics, TU Dortmund University, Dortmund, Germany 
\and
Matthew Wheeler \textit{matt.wheeler@nih.gov} \\
Biostatistics and Computational Biology Branch, National Institute of Environmental \\ Health Sciences, Research Triangle Park, Durham, North Carolina, U.S.A. }

\date{\today}

\maketitle

\begin{abstract}
When modeling biological responses using Bayesian non-parametric regression, prior information may
be available on the shape of the response in the form of non-linear function spaces that define the general shape of
the response. To incorporate such information into the analysis, we develop a non-linear functional shrinkage (NLFS)
approach that uniformly shrinks the non-parametric fitted function into a non-linear function space while allowing for
fits outside of this space when the data suggest alternative shapes. This approach extends existing functional
shrinkage approaches into linear subspaces to shrinkage into non-linear function spaces using a Taylor series expansion
and corresponding updating of non-linear parameters. We demonstrate this general approach on the Hill model, a
popular, biologically motivated model, and show that shrinkage into combined function spaces, i.e., where one has
two or more non-linear functions a priori, is straightforward. We demonstrate this approach through synthetic and
real data. Computational details on the underlying MCMC sampling are provided with data and analysis available
in an online supplement.
\end{abstract}


\section{Introduction}
\label{s:intro}

When modeling complex biological systems, mechanistic knowledge about the system under investigation is often available; however, including this information in a statistical model may be impossible due to the system's complexity in relation to experimental and computational resources  \citep{mesarovic2004search}.
Often, simplified models are used in lieu of the true mechanistic model \citep{vsimon2005considerations}.
When using these simplified models, one expects them to describe the observed data correctly or be mildly misspecified, and in the case of misspecification, the model may still be helpful in describing the response. 

When modeling biological systems, an example of this situation is the use of the Hill model.  This model, which represents sigmoidal-shaped responses,  is a simplification of the complex biochemical process based upon chemical kinetics \citep{hill1910possible} and is used to model a wide variety of biochemical processes \citep{goutelle2008hill}. Despite its widespread use, it may not always represent the observed response. Non-monotone deviations of the Hill’s functional form may be evident in the data.  Additionally, other competing models may also be available, and the modeler might like to include this information to inform the fitting process, too. We develop a framework that allows one to define a subspace over one or more function spaces of interest for Bayesian non-parametric regression.  

From the Bayesian perspective, there is a rich literature on approaches incorporating prior knowledge in non-parametric regression. Naively, one may center the non-parametric model on the specified parametric function. 
When the parametric data-generating mechanism's mean is the known parametric model, ensuring that estimates do not contain artifactual deviations from that model is difficult, implying that shrinkage to the prior model will not be uniform. 
Further, using this method, there is no way to create a space based on multiple parametric functions. More sophisticated approaches use shape constraints induced through the prior distribution, which include monotonicity or limits to the number of extrema \citep{brezger2008, shivley2009, mayer2008, shively2011, meyer2011, gunn2005transformation, kollmann2014unimodal, wheeler2017bayesian}. Though these approaches are often effective, they do not directly incorporate parametric modeling information on the shape of the model; they force the response to be in the constrained space by putting a prior mass of zero on all responses outside of that space.

Alternatively, one may merge mechanistic prior knowledge into a model is through ordinary differential equations (ODEs) within a Bayesian framework. Parametric Bayesian models include pharmacokinetic/pharmacodynamic modeling, discussed by \cite{lunn2002bayesian}, and \citet{huang2006hierarchical} present an HIV-modeling example using Bayesian hierarchical models with non-linear differential equations.
 More flexible non-parametric approaches use differential equations to inform stochastic processes with induced constraints \citep{golightly2011bayesian, titsias2012identifying}. While 
 \citet{alvarez2013linear} and \citet{wheeler2014mechanistic} proposed a Gaussian Process (GP) approach that incorporates mechanistic knowledge defined by differential equations. More recently, \cite{chen2022apik} incorporate mechanistic knowledge defined by linear or non-linear partial differential equations (PDE) into a GP framework by selecting PDE points, i.e. pseudo covariate points through which the assumed PDE information is incorporated.
 Like the shape-constrained approaches, these methods form a Basis expansion consistent with a subspace defined using mechanistic knowledge. Thus, these priors imply that an estimated function is within the given subspace, and they do not allow for deviations outside of this space. 

 We define a prior distribution over a non-linear subspace  - such as the Hill model and power model - that does not require a fitted function to be within that subspace.  When the non-linear subspace is correctly specified, shrinkage into it occurs; but, when the true model is outside of the subspace, the approach is unconstrained. 
 We build upon the work of \citet{shin2020functional} who introduced the functional horseshoe (fHS) prior for linear spaces. The fHS prior shrinks the non-parametric fit towards a pre-specified, linear subspace. This approach is different from well-known shrinkage approaches such as Ridge, Lasso or Horseshoe \citep{hoerl1970ridge, tibshirani1996regression, carvalho2010horseshoe}, which shrink model coefficients in a non-parametric regression towards the origin. The prior of \citet{shin2020functional} has the appealing property that the posterior shrinks into the pre-specified subspace f it is consistent with the observed data or, alternatively, is left unconstrained otherwise.
 The shrinkage occurs at the minimax optimal rate.

In our extension, we use a Taylor expansion to locally linearize the response function, where the derivatives depend on parameters of the non-linear model.  The extension allows functional shrinkage into a non-linear function space or adapts the function to be outside of the non-linear space.  
The relevant non-linear function space is specified \textit{a priori} using one or more parametric models.

We present our shrinking approach in Section \ref{s:model}. Section \ref{s:single_combined} then illustrates the approach both for the case of shrinkage into a single function space - shown for the Hill model - and into a combined function space  - shown for the Hill and the power models.
We compare our method against other parametric and non-parametric approaches in a simulation study in Section \ref{s:simulation}.
We apply our method to a real-world data example of total testosterone levels measured in 9943 males aged between 3 and 85 years in section \ref{s:real_data}. The computational back-bone of the approach is MCMC sampling combining Gibbs-, Metropolis-Hastings- and Slice-sampling \citep{brooks2011handbook, neal2003slice}, detailed in the supplementary material.

\section{Model}
\label{s:model}

\subsection{Spline Model}

Consider the non-parametric regression problem
\begin{align}
    y_i = g(x_i) + \varepsilon_i, 
\end{align}
with unknown mean function $g: \mathbb{R} \rightarrow \mathbb{R}.$ 
We observe  $y = (y_1, \dots, y_n)'$ corresponding to covariates $x = (x_1, \dots, x_n)' $ and wish to estimate $g.$ Assuming $\varepsilon_i  \overset{\text{iid}}{\sim} N(0, \sigma^2),$ it is common to approximate $g$ using a B-spline basis expansion \citep{carl2001practical}, i.e., 
\begin{align}
    f(x_i) = \sum_{m=1}^{k} \phi^{j}_{m}(x_i) \beta_m \label{sec2:spline},
\end{align}
or $f(x) = \Phi \beta$. Here, the B-spline basis $\phi_k^j(x)$ are of order $j$  defined on $k^*$ internal knots, where $k = k^*+j,$  $\beta = (\beta_1, \dots, \beta_k)^{\top}$ denotes the vector of basis coefficients.  We consider cubic splines and omit the superscript $j = 3$.
With a dense knot set,  the spline approximation $f$ can be made to be arbitrarily close to any continuous $g,$  allowing one to estimate a large space of functions to arbitrary precision.

\subsection{Bayesian Priors for a Non-linear Subspace}
For many prior specifications, the expansion in (\ref{sec2:spline}) may not place high prior probability on  biologically relevant responses. To define a biologically relevant model, we construct a prior distribution that places significant prior mass on the function space defined by the non-linear model, e.g., the space of Hill models, but does not put zero mass outside the function space.

To do this, assume knowledge about the shape of $g$ through a twice differentiable function $h_\theta: \mathbb{R} \rightarrow \mathbb{R}$.
The function $h_\theta$ depends on parameter vector $\theta,$ and defines the function space $\Omega^{\Theta}_0 = \{h_{\theta} | \theta \in \Theta\}$ for all realizations $\Theta \subseteq \mathbb{R}^s$.
If the true mean function $g$ happens to be outside $\Omega^{\Theta}_0$, shrinkage towards $\Omega^{\Theta}_0$ is undesirable. Given a dense knot set,  the spline $f$ can approximate $h_{\theta}$ for any $\epsilon-$ball. Consequently, the space of functions represented by the spline contains $\Omega^{\Theta}_0.$  We define a prior for (\ref{sec2:spline}) that places prior mass on $\Omega^{\Theta}_0,$ but does not limit responses to be only in $\Omega^{\Theta}_0$.

To define this prior, we consider \citet{shin2020functional},
who defined a projection prior that shrinks into the linear column space defined by the matrix $\Phi_0 \in \mathbb{R}^{n \times d}$ through
\begin{equation}
\label{eq:shin_proj}
        p(\beta | \sigma^2, \tau^2) \propto (\tau^2)^{-(k - d_0)/2} \exp \left( - \frac{1}{2 \sigma^2 \tau^2} \beta^{\top} \Phi^{\top} (I - P_{\Phi_0}) \Phi \beta \right),
\end{equation}
where $d_0 = \text{rank}(\Phi_0),$ $\Phi_0$ is constructed as a linear space of known covariates, and $P_{\Phi_0}$ is the orthogonal projection matrix into the column space of $\Phi_0.$
The hyperparameter $,\tau,$ is given a generalized horseshoe (HS) prior with hyperparameters $a$ and $b$ (cf. \cite{shin2020functional}). When $a=b=0.5$ the prior is a half-Cauchy distribution, and one arrives at the HS prior \citep{carvalho2010horseshoe}.

In (\ref{eq:shin_proj}),  one constructs $P_{\Phi_0}$ using the linear column space of $\Phi_0.$ Given our space is non-linear, there is no direct analogue to $P_{\Phi_0}.$  As an approximation, we use a Taylor series approximation of $h_{\theta_0},$ $\theta_0 \in \Theta.$  That is, we linearly approximate $\Omega^{\Theta}_0$ at any $\theta_0$ using a first-order Taylor expansion
\begin{equation}
    h_{\theta}(x) \approx h_{\theta_0}(x) + \dot{\text{H}}_{\theta_0}(x)(\theta-\theta_0)
\end{equation}
where $\dot{\text{H}}_{\theta_0}(x) = \left. \frac{\partial h_{\theta}(x)}{\partial \theta} \right\rvert_{\theta = \theta_0}  \in \mathbb{R}^{n \times s}$ is the Jacobian containing the partial derivatives of $h_{\theta}$ evaluated at $\theta_0$.
The column space of $\dot{\text{H}}_{\theta}(x)$ approximates $h_{\theta}(x)$  \citep{seber2003nonlinear}[p. 130] and we use $\dot{\text{H}}_{\theta}(x)$ to construct  $P_{\theta} = P_{\dot{\text{H}}_{\theta}}$.
Thus, for any $\theta_0$, we project $f(x)$ onto the  space locally approximating $h_{\theta_0}.$ When there are multiple function spaces to consider, the same approach applies;
in this case, operator $P_{\dot{\text{H}}_{\theta}}$ defines the projection into a combined linear space, where $\dot{\text{H}}_{\theta}$ represents the Jacobian across all assumed functions. 

We place the prior 
\begin{equation}
    \label{eq:beta_prior}
        p(\beta | \sigma^2, \tau^2, \theta) \propto (\tau^2)^{-k/2} \exp \left( - \frac{1}{2 \sigma^2 \tau^2} \beta^{\top} \Phi^{\top} (I - P_{\theta}) \Phi \beta \right)
\end{equation}
over $\beta$ to shrink realizations of (\ref{sec2:spline}) into $\Omega^{\Theta}_{0}.$ In (\ref{eq:beta_prior}), $\theta$ is given an appropriate prior to complete the specification. This approach penalizes deviations of $\Phi \beta$ based upon the projection operator $(I-P_{\theta}).$ As we shrink back to a planar approximation given a specific $\theta_0,$ we require appropriately specified priors for the non-linear parameters in $\Theta$. As $\beta$ is defined conditional on $\theta$ through the
linear projection operator $P_{\theta},$ only priors for the non-linear parameters can be learned. 

In this formulation, $(\tau^2)^{-(k-d_0)/2}$ in (\ref{eq:shin_proj}) becomes $(\tau^2)^{-k/2}$ because we separately model the intercept (cf. Section \ref{s:hill_case}).  This change yields proper priors as due to the non-linearity, no linear basis of $\Phi$ is in $(I-P_{\theta})$ and $\Phi^{\top}(I-P_{\theta})\Phi$ has full rank. 

\section{Non-linear functional shrinkage for single or combined function spaces}
\label{s:single_combined}

\subsection{Single function spaces}
\label{s:hill_case}

As an example of non-linear functional shrinkage using a single function, we consider the Hill model.  This function is given by
\begin{equation}
    h(x, \theta) =  \theta_1 + \theta_2 \frac{x^{\theta_4}}{{\theta_3}^{\theta_4} + x^{\theta_4}}, 
    \label{eq:hill}
\end{equation}
where $\theta_1$ is the background response at $x=0$, $\theta_2$ is the maximal change in the response, $\theta_3$ is the dose where half of this change is reached and $\theta_4$ defines the steepness of the curve.
The Jacobian,  $\dot{\text{H}}_{\theta},$ is 
\begin{equation}
\label{eq:deriv}
    \left. \frac{\partial h(x, \theta)}{\partial \theta} \right\rvert_{\theta = \theta_0}  = \left. \begin{pmatrix} 
    1 & 
    \frac{x^{\theta_4}}{x^{\theta_4}+ {\theta_3}^{\theta_4}} &
    \theta_2 \frac{-\theta_4}{\theta_3} s(x, \theta_3, \theta_4) &
    \theta_2 \log(\theta_3 / x) s(x, \theta_3, \theta_4)
    \end{pmatrix} \right\rvert_{\theta = \theta_0},
\end{equation}
with $s(x, \theta_3, \theta_4) = ( ( x{\theta_3}^{-1})^{\theta_4} +1)^{-1} (( \theta_3 x^{-1})^{\theta_4} +1)^{-1}$. 
The derivative matrix does not depend upon the linear parameters $\theta_1$, but it still depends on $\theta_2$. However, $P_{\theta}$ does not depend on $\theta_1$ and $\theta_2$ (cf. Lemma 1 in the appendix), which gives a direct example of why we do not place a prior over these linear quantities.
Of the parameters in (\ref{eq:deriv}), parameter $\theta_3$ is of particular interest because it represents the value of $x$ that produces a response that is the average of the lower and upper asymptote. Values of the covariate below $\theta_3$ correspond to values of the response less than $50\%$ of the maximal response.  Further, $\theta_4$ corresponds to the steepness of the response and speed of a chemical reaction in a biological substrate.  As both quantities have direct interpretation, informative priors can be developed for these quantities accordingly, which in turn informs the subspace the model may shrink into.

To specify the hyperprior over $(\theta_3, \theta_4)$, we assume $x \in [0, 1],$ and let $E[\theta_3] = 0.5,$  the midpoint, and for $\theta_4,$ we center it on $3$, letting the parameter vary within a range that we have often seen in bioassays. In our model, $\theta_1$ enters as the intercept, and $\theta_2$, the maximal response change, implicitly enters the model through the 
$\beta$ coefficients. Using the Hill model as a prior to define (\ref{eq:beta_prior}), we complete the prior specification as
\begin{align}
\label{eq:model_interc}
    (y|\beta, \sigma^2, \theta_1) &\sim N(\theta_1 + \Phi \beta, \sigma^2 I_n) \\
    \theta_1 \sim N(0, 20), \quad
    \theta_3 &\sim N_{+}(0.5, 0.05) \quad
    \theta_4 \sim LN(0.95, 0.29) \\
    \sigma^2 &\sim IG(0.001, 0.001),
\end{align}
where $N_{+}(a,b)$ is a truncated normal distribution with mean $a$ and variance $b$ (before truncation), $LN(a,b)$ is a log-normal distribution with log-mean $a$ and log-variance $b,$ and $IG(a,b)$ is an inverse-gamma distribution with shape $a$ and scale $b$. Note that $\theta_4 \sim LN(0.95, 0.29)$ results in $E[\theta_4]=3$ and $V[\theta_4] = 3$.

\subsection{Combined function spaces}
\label{s:hill_power_combined}

If one desires multiple functions to define in the function space because of uncertainty in the  function space, one can add multiple functions.   Here, assume there are $r \in \{1, \dots, R\} = \mathcal{R}$ function spaces $\Omega_{0}^{(r)} = \{h_{\theta}^{(r)} | \theta \in \Theta^{(r)}\}$ of interest;  we omit the index $r$ on each $\theta$ for simplicity.
For each $\Omega_{0}^{(r)},$ calculate the Jacobian, $\dot{\text{H}}_{\theta}^{(r)},$ i.e.,
\begin{equation*}
    \dot{\text{H}}_{\theta}^{(\mathcal{R})} = (\dot{\text{H}}_{\theta}^{(1)} \hdots \dot{\text{H}}_{\theta}^{(R)}),
\end{equation*}
and use this to construct $P_{\theta}.$ 
The Jacobian, $\dot{\text{H}}_{\theta}^{(\mathcal{R})},$ must be full rank without linear bases other than an intercept column for Equ. \ref{eq:beta_prior} to hold.

To illustrate the combined subspace shrinkage approach, we use the Hill and power models.  The latter function defined as as $h_{\theta}(x) = \theta_1 + \theta_2x^{\theta_3},$ which has only one non-linear parameter, $\theta_3,$ that requires a prior specification.
We use $\theta_3 \sim N(0.5, 0.25)$, to center on a concave shape.
The partial derivatives of the power model are \begin{equation}
    \frac{\partial h(x, \theta)}{\partial \theta} =\left(1 \quad x^{\theta_3 } \quad \log(x)x^{\theta_3} \right) = \dot{\text{H}}_{\theta}^{(1)}.
\end{equation}
Prior to combining $\dot{\text{H}}_{\theta}^{(1)}$ of the power model and $\dot{\text{H}}_{\theta}^{(2)}$ of the Hill model (Eq. \ref{eq:deriv}), we remove the intercept from $\dot{\text{H}}_{\theta}^{(1)}$ to obtain a full rank. Shrinkage into the combined subspaces is equivalent to shrinkage into a single subspace.

\section{Simulation Study}
\label{s:simulation}

\subsection{Setup}
\label{s:simulation_setup}
We perform a simulation study and evaluate the performance of the proposed approach against other fitting strategies. Full details of the simulation design are summarized by the ADEMP principle described in \citet{morris2019using} (Table S\ref{tab:sim_setup}). We generate data using three parametric cases: the Hill model, the power model, and a misspecified model (the Hill model with downturn). We look at exposure-response data as, for such data,  chemical kinetics of exposure are  often approximated by the Hill model, but the results generalize to other domains. 

For each data set, we draw $x \in [0, 1]$ uniformly for $n \in \{50, 100, 200, 500\}$ observations, where $50$ is a realistic assay size and larger $n$ are chosen to study the large sample behavior.
Mean zero normal noise with variance $\sigma^2 = 0.005$ and a larger noise $\sigma^2 = 0.05$ is added.
These variances represent a 2-SD spread that is approximately $14\%$ and $45\% $ of the maximal response. In total, $24$ data generating scenarios are used, with $n_{\text{rep}} = 1000$  simulations per scenario. For each dataset, we apply the following methods: 

\subsubsection{Modeling Approaches} 

\subsubsection*{Non-linear functional shrinkage (NLFS)\\}

Non-linear functional shrinkage is performed with shrinkage into the Hill space (NLFS(Hill)), power space (NLFS(power)), or a combination (cf. Section \ref{s:hill_power_combined}) of both function spaces (NLFS(Hill+power)). Two variations for the shrinkage parameter $\tau^2$ are considered. One uses a half Cauchy prior ($a=b=0.5$) and is implemented according to \cite{makalic2015simple}; the other, implemented ourselves using slice sampling \citep{neal2003slice}, uses a $\omega \sim \text{Beta}(a, b)$ prior where $a=0.5$ and $b=\exp(-k\log(n)/2)$ as proposed by \cite{shin2020functional} and $k$ is the number of knots.
    
\subsubsection*{ Parametric Model (Param.)\\} 

We investigate the performance fitting of the Oracle model using Bayesian parametric regression for the Hill model (Param.(Hill)) or the power model (Param.(power)) (priors in Table S\ref{tab:sim_setup}). Fitting these models allows us to compare the performance of the Oracle NLFS to the Oracle parametric model. 
 
\subsubsection*{B-splines\\} 

Bayesian B-splines with a scaling parameter $\lambda^2\sim\text{IG}(0.001, 0.001)$ where the spline coefficients are given by the prior $\beta \sim N(0, \sigma^2 \lambda^2 \text{diag}(k)).$ This model represents a basis approach without smoothing and is used to compare the performance of the NLFS approach when the shrinkage subspace is misspecified. 

\subsubsection*{ P-splines\\} 

Penalized Bayesian smoothing splines where $\beta \sim N(0, \sigma^2 \tau^2 K^{-1})$ where $K= R^{\top} R$ and $R$ is a second order penalty matrix and $\tau^2 \sim \text{IG}(1, 0.005)$, similar to the hyperparameter choices in \cite{lang2004}. This approach builds upon the B-spline approach, adding a smoothing component, and typically performs better in practice than B-splines

\subsubsection*{Parametric Model + horseshoe B-spline \\}

We also consider a model that includes the true parametric model plus an additional B-spline to account for model misspecification, i.e.,  $y = h_{\theta}(x) + \Phi \beta + \varepsilon.$  
When $h_{\theta}(x)$ specifies the correct model, one has $\beta = 0$; otherwise, $\beta \neq 0$.  To shrink the $\beta$ coefficients to zero, we use a horseshoe prior, i.e.,  $\beta \sim N(0, \sigma^2 \tau^2 \text{diag}({\lambda_1}^2, \dots, {\lambda_k}^2))$ where $\tau \sim C^{+}(0, 1)$ and $\lambda_j \overset{\text{iid}}{\sim} C^{+}(0, 1)$, cf. \cite{makalic2015simple}.
$C^{+}(0,1)$ denotes a standard Half-Cauchy prior.
As in the parametric model case, $h_{\theta}$ is either the parametric Hill (Param.(Hill)+B-spline) or power model (Param.(power)+B-spline). This approach represents a direct competitor to the NLFS approach.

\subsubsection{Further Considerations\\}

For all simulations, we use $k=15$ inner knots for the B-spline basis matrix. When MCMC sampling, we took 10,000 draws from the posterior, discarding the first 2000 samples as burn-in. Initial experiments indicated that this  number of samples was adequate to estimate the posterior distribution.  For the spline-based approaches (NLFS, B-spline, P-spline), we place a vague prior on the intercept term, $\theta_1,$ defined in (\ref{eq:model_interc}).
Traceplots, of an NLFS fit with correctly and incorrectly specified subspaces, are given in the supplement (supplemental Figures \ref{fig:traceplots_correct} and \ref{fig:traceplots_misspec}) and show convergence.

\subsection{Results}
\label{s:results}

Figure \ref{fig_sim_res_red} gives representative results of the simulation, where all results are provided in the supplement. 
Unsurprisingly, when the Hill model is the truth (Figure \ref{fig_sim_res_red}a), we observe the largest RMSE of 0.151 when fitting the misspecified parametric model  (Param.(power)). Unlike the misspecified parametric fits, when the function space is misspecified in the NLFS approach (NLFS(power)), the RMSE is approximately one-third (0.046) that from fitting the misspecified parametric model, indicating the NLFS approach adjusts to the data.  In this scenario,  the NLFS(power) performance with mean RMSE of 0.046 was similar to that of the B-spline approach with mean RMSE of 0.042.

\begin{figure}
    \centering
    \includegraphics[width = 0.95\linewidth]{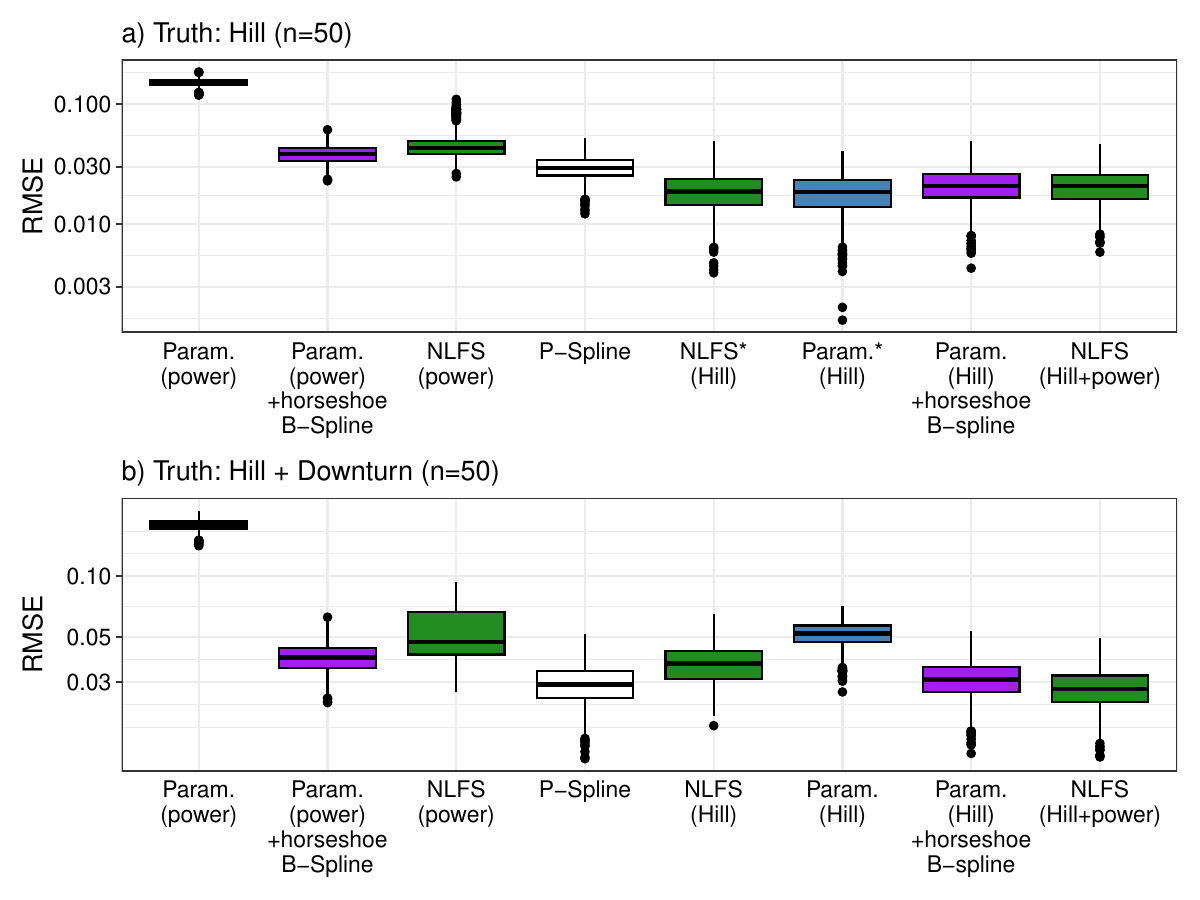}
    \caption{
    Representative root mean square error (RMSE) results of the simulation for the scenarios where the truth is Hill (a) or Hill + Downturn (b) and a medium noise level. Pane b represents the situation where a deviation of an unknown shape is the truth. All simulation results can be found in Tables S\ref{tab:sim_res_sigma_sq_0.0005} and S\ref{tab:sim_res_sigma_sq_0.005}.
    $^\ast$Correct model used in the model fit.    
    }
    \label{fig_sim_res_red}
\end{figure}

When the correct function space is assumed for the NLFS prior, the RMSE drops to 0.019 (Figure \ref{fig_sim_res_red}a), as low as that of the 0racle parametric fit.
This demonstrates the adaptive shrinkage behavior of the NLFS approach in the case of correct subspace specification. Here, the NLFS approach effectively shrank towards the correctly assumed space for sample sizes as low as $n=50$, and performed similarly to the oracle parametric model fit. The P-spline approach receives no prior model or subspace specification but yields smooth splines.  Consequently, its performance was in between the approach with misspecification and correct specification.

When the correct function space is assumed, the NLFS approach tended to outperform the parametric + horseshoe B-spline (PHBspline) approach.
The PHBspline approach does not enforce an equally smooth, global shrinkage of all $\beta$ towards zero, especially when there are leverage points far from the observed mean.  

When all assumed models or spaces are misspecified (Figure \ref{fig_sim_res_red}b),  the NLFS approach was outperformed by the PHBspline approach for the same model misspecification, i.e., NLFS(power) was outperformed by param.(power) + horseshoe B-spline and NLFS(Hill) was outperformed by param.(Hill) + horseshoe B-spline.
However, the NLFS approach in general has an advantage in misspecification scenarios due to its inherent flexibility to shrink toward combined function spaces.
The NLFS(Hill+power) outperformed all other approaches in this scenario with a mean RMSE of 0.028.
Only the P-spline approach came close, showing a slightly weaker performance with a mean RMSE of 0.030.

\begin{figure}
\centering
\begin{subfigure}{0.49\textwidth}
    \includegraphics[width=\textwidth]{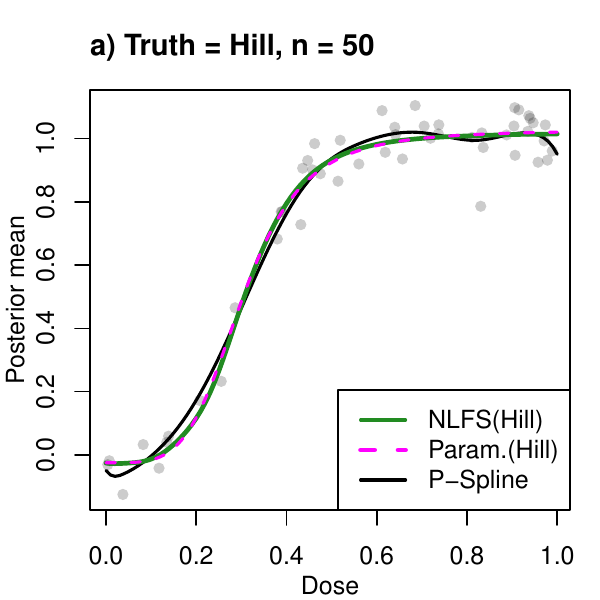}

    \label{fig:ex_hill}
\end{subfigure}
\hfill
\begin{subfigure}{0.49\textwidth}
    \includegraphics[width=\textwidth]{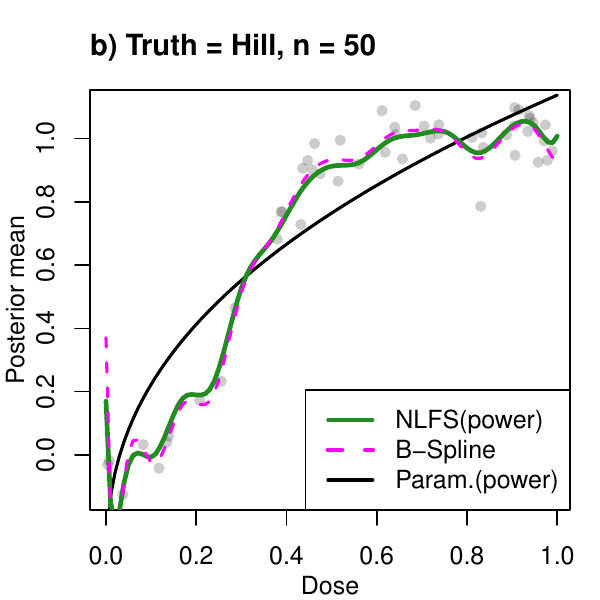}

    \label{fig:ex_misspec}
\end{subfigure}

\begin{subfigure}{0.49\textwidth}
    \includegraphics[width=\textwidth]{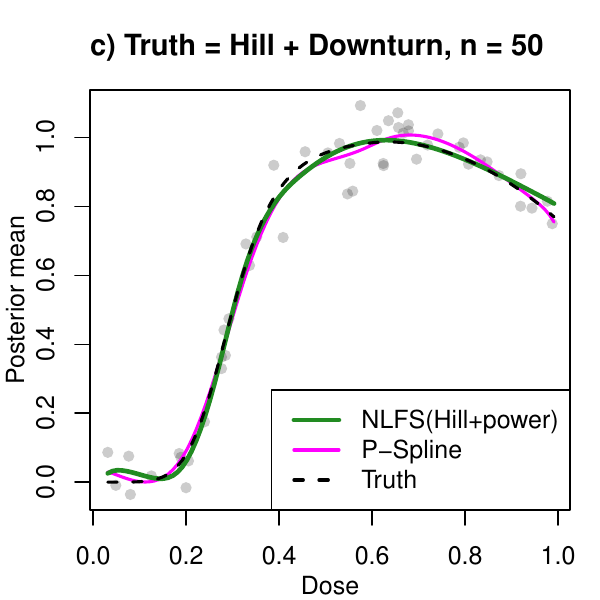}

    \label{fig:ex_hilldown}
\end{subfigure}
\hfill
\begin{subfigure}{0.49\textwidth}
    \includegraphics[width=\textwidth]{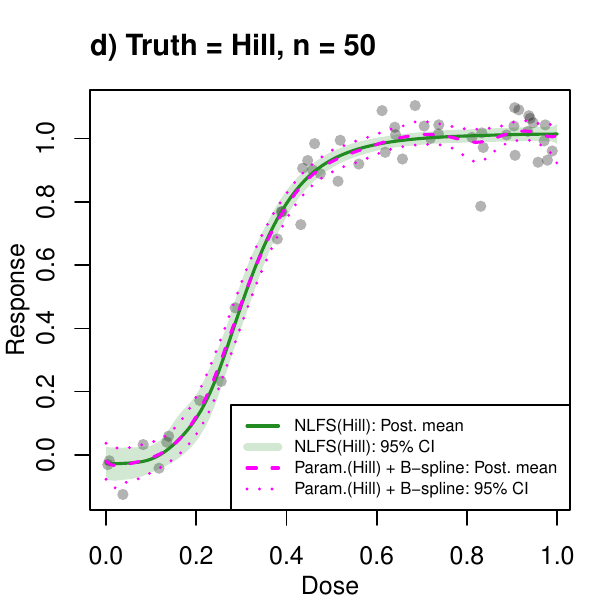}

    \label{fig:ex_CI}
\end{subfigure}
        
\caption{Posterior mean responses (a-d) and credible intervals (d) of representative simulation runs with median noise level ($\sigma^2=0.005$): (a) Oracle scenario. (b) misspecification scenario. (c) deviations of unknown shape. Combined subspace shrinkage  reduces misspecification errors. (d) Comparison of credible intervals between NLFS and PHBspline in the Oracle scenario.}
\label{fig:examples}
\end{figure}

The NLFS prior appropriately shrinks into the correct space, giving equivalent fits to the parametric Hill model (Figure \ref{fig:examples}a). Correspondingly, the P-Spline smoothing approach estimate shows various artifactual bumps not evident in the NLFS approach.  
Even if the true model is the Hill model, the naive PHBspline approach produces an artifactual bump in the asymptote region that does not occur with the NLFS fit (Figure \ref{fig:examples}d). 
The NLFS approach and generic B-spline are equivalent when the model is misspecified (i.e., fits a model outside of the assumed space) (Figure \ref{fig:examples}b).
The NLFS(Hill+power) approach fits a model outside of the Hill space (Figure \ref{fig:examples}c) and illustrates how shrinkage into combined subspaces can reduce misspecification errors involving minor deviations.

\section{Real Data Example}
\label{s:real_data}

We applied the proposed method on a testosterone data set collected by \cite{kelsey2014validated}.
They modeled total testosterone (TT) concentration in male participants dependent on age, to identify normal TT ranges at any age.  TT levels are the result of highly complex physiological processes, mechanistic models are not available. TT is expected to increase during puberty, reach a maximum and possibly slowly decline with age, which is a sigmoidal assumption.   Consequently, we \textit{a priori} assume the Hill based shape through the NLFS prior.

\cite{kelsey2014validated} collected data from 13 studies on TT by age, yielding $>$10.000 data points; they then fit 330 polynomial models and selected a single best parametric model based on the best $R^2$ with 5-fold cross-validation. Due to the large number of data points, spline smoothing approaches tend to produce local artifacts that are biologically unreasonable. The NLFS  approach offers an alternative to extensive model comparisons while simultaneously incorporating knowledge on the curves shape.

For the analysis,  we set
\begin{align}
    \theta_3 &\sim N(15, 4), \\
    \theta_4 &\sim LN(2.28, 0.05).
\end{align}
Since testosterone levels increase during puberty, \textit{a priori} we assume TT levels to reach half maximal levels (e.g., $\theta_3$) at approximately 15 years with a standard deviation of 2 years. For the steepness parameter, $\theta_4,$  we set the log-mean to 2.28 and the log-variance to 0.05 implying that the actual mean and variance to be 10 and 5, respectively. We choose these values as we expect TT to increase rapidly upon the onset of puberty. 

We fit our model to the slightly reduced data set of men of age 85 or younger (98.5\% of original data) because of extreme variability in the approximately 150 observations above 85.
To model the noise, we let  $\sigma^2 \sim C^{+}(0, 1)$, to account for the large variance in the data. 

\begin{figure}
    \centering

    \includegraphics[width=0.6\linewidth]{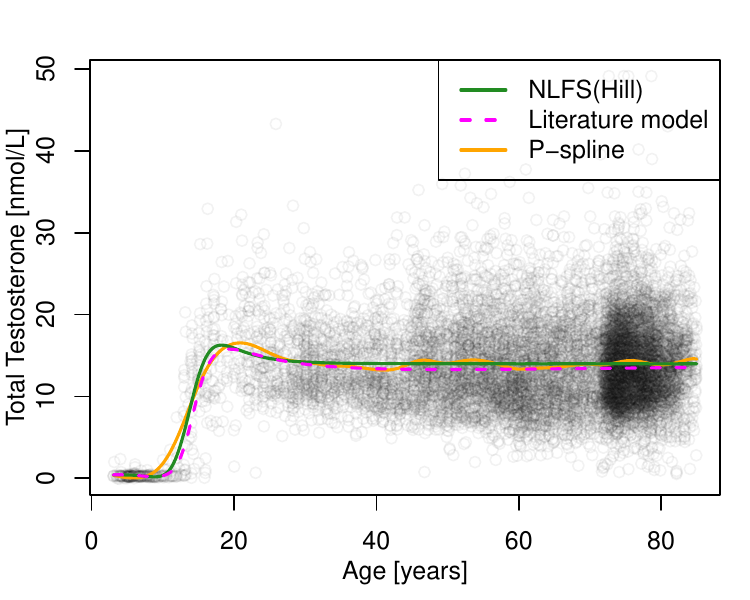}

    \caption{Comparison of the NLFS approach to a P-spline fit and the best parametric model selected by \cite{kelsey2014validated}. For the P-spline and NLFS(Hill), the posterior mean is displayed. The parametric model by Kelsey is $\log_{10}(TT+1) = (a + cx + ex^2 + gx^3) / (1+bx+dx^2+fx^2)$ with $a=0.04655$, $b=-0.05311$, $c=0.05123$, $-d=0.00793$, $e=-0.01222$, $f=0.00058$ and $g=0.00069$. For display, the model is backtransformed using $\exp(\log_{10}(TT+1) + \hat{\sigma}^2 / 2)$ where $\hat{\sigma}^2$ is the estimated residual variance.}
    \label{fig:testosterone}
\end{figure}

The NLFS and parametric fit by \cite{kelsey2014validated} are roughly sigmoidal (Figure \ref{fig:testosterone}).  They both show a peak around age 19, followed by a slight descent that eventually plateaus. The NLFS
approach notably predicts a larger mean testosterone level than the parametric model. The P-spline fit has a less pronounced peak TT around age 20 and has artifactual bumps that oscillate around the NLFS estimate.
For each method, the observed RMSE values were 5.123 (NLFS), 5.154 (parametric), and 5.111 (P-spline) and therefore similar, but the three resulting model fits are visibly distinguishable. Arguably, the NLFS approach more appropriately models the mean than the P-spline due to the latter’s bumpiness. 
Further, NLFS estimates a higher mean TT level than the parametric model, which suggests there may be some underestimation of the mean TT when using a parametric approach. 

\section{Discussion}
\label{s:discussion}

The NLFS prior enables adaptive shrinkage into a pre-specified non-linear function space but does not constrain the resulting function to be in that space if the data are not compatible with that \textit{a priori} function space. This approach can be applied to function spaces defined by any twice differentiable function. Because such a setting is common, the NLFS approach balances adhering to prior assumptions and accounting for model misspecification.

The NLFS approach can shrink into a combined function space, thereby providing robustness against misspecification. This benefit is supported by simulation results, where NLFS combined function space prior outperformed all other methods under model misspecification. Defining such a prior is straightforward for the NLFS approach. Attempting to account for parametric misspecification by including a spline that shrinks to zero if the model is correctly specified can give artifactual features in the Oracle scenario. As a comparison, the NLFS prior gave slightly better RMSE results under the Oracle scenario, and provided more realistic curve fits without the artifactual features. Though NLFS with a misspecified single subspace performed slightly worse, adding subspaces in NLFS did not lead to a relevant performance loss compared to only including the correct model but also robustified NLFS against misspecification.

When modeling the TT data in  \cite{kelsey2014validated}, the 
NLFS yielded a plausible non-parametric estimate that did not produce artifactual features.  In this regard, NLFS provided equally reasonable mean estimates as the parametric model, while not requiring a model selection procedure on over 300 models.

Simulation results empirically show that NLFS correctly decides to either shrink towards the specified subspace, or remain unconstrained. Though we have not provided a theoretical proof, our simulation results suggest that the optimal, theoretical shrinkage properties given by \citet{shin2020functional} approximately hold in the non-linear case. Because it performs similarly to an unsmoothed spline estimate, adding a smoothness penalty similar to the one proposed by \cite{wiemann2021adaptive} for linear subspace shrinkage may be a promising extension.

The construction of NLFS assumes the independent variable to be continuously distributed, with a unique covariate value for each observation.  In some biological applications, data are generated in a planned experimental setting, with multiple units treated at few distinct exposure levels. For such experiments, the exposure is typically a dose or concentration. Dose-response modeling is often performed in terms of a simple parametric Hill model fit, which can lead to misspecification errors that could be prevented using NLFS. Tailoring NLFS to a such a data structure is necessary. This can be done using a grid that defines the shrinkage locations, such that the shrinkage is independent of the few experimentally selected doses.
Precisely, $\Phi_{\theta}$ would  be evaluated at a grid instead of the observed exposure levels. This avoids a lack of shrinkage at basis functions that fall between exposure levels.
This extension would yield more model parameters related to the construction of the grid. Another extension using a fully specified Gaussian processes is an alternative and would reduce hyperparameter choices on knot sequences and shrinkage grids.
Another extension is to account for heteroscedasticity. 
For non-parametric Bayesian modeling, different methodologies can be applied, e.g. Dirichlet process priors. 
Other computational challenges in the NLFS approach relate to the derivatives. For example, using the Hill model, derivatives w.r.t. the non-linear parameters can be almost linearly dependent.
Careful prior selection or expanding the shrinkage onto additional subspaces might soften this challenge.

\newpage

\appendix

\section*{Acknowledgements}

This manuscript was funded in part by the Research Training Group “Biostatistical Methods for High-Dimensional Data in Toxicology” (RTG 2624)  funded by the Deutsche Forschungsgemeinschaft (DFG, German Research Foundation-Project Number 427806116) and intramural funds at the NIEHS. 

\section*{Supporting Information}

The code and data underlying this article are available on GitLab at \newline \href{https://gitlab.tu-dortmund.de/functional_shrinkage/nonlinear_shrinkage}{https://gitlab.tu-dortmund.de/functional\_shrinkage/nonlinear\_shrinkage}.


\section{Tables}

\begin{table}[ht]
\centering
\caption{Overview of method settings used in the simulation study.}
\label{tab:sim_methods}
\begin{tabular}{rlll}
  \hline
 & Algorithm & Assume & Shrinkage (Horseshoe prior) \\ 
  \hline
1 & NLFS & Hill & $a=b=0.5$ \\ 
  2 & NLFS & Hill & $a=0.5, \, b=\exp(-k\log(n)/2)$ \\ 
  3 & NLFS & Hill \& power & $a=b=0.5$ \\ 
  4 & NLFS & Hill \& power & $a=0.5, \, b=\exp(-k\log(n)/2)$ \\ 
  5 & NLFS & power & $a=b=0.5$ \\ 
  6 & NLFS & power & $a=0.5,\, b=\exp(-k\log(n)/2)$ \\ 
  7 & parametric & Hill &  -\\ 
  8 & parametric & power &  -\\ 
  9 & B-spline & - &  -\\ 
  10 & P-spline & - &  -\\ 
  11 & Parametric + B-spline & Hill & $a=b=0.5$ \\ 
  12 & Parametric + B-spline & power & $a=b=0.5$ \\ 
   \hline
\end{tabular}
\end{table}

\begin{table}[htb]
    \caption{Simulation setup summarized by the ADEMP principle.}
    \label{tab:sim_setup}
    \centering
    \small
    \begin{tabular}{l|p{13cm}}
    \hline
        \textbf{A}im & Comparing proposed approach against existing approaches  \\
        
        \textbf{D}ata generation & Dose-response models:\\
         & - Hill: $h_{\theta}(x) = \frac{x^{\theta_4}}{\theta_3^{\theta_4} + x^{\theta_4}}$ ($\theta_3 = 0.3, \theta_4 = 6$) \\
         & - Power: $h_{\theta}(x)=\theta_1x^{\theta_2}$, ($\theta_2 = 0.5$)\\
         & - Hill + Downturn: $h_{\theta}(x) = h_{\theta}^{\text{Hill}}(x) + \mathbbm{1}_{[0.6, \infty)}(x)(-1.5(x-0.6)^2)$ ($\theta_3 = 0.3, \theta_4 = 6$) \\
        & Doses: Unif$\sim [0, 1]$ \\
        & Sample sizes: $n \in \{50, 100, 200, 500\}$ \\
        & Added noise: $\varepsilon \sim N(0, \sigma^2)$, $\sigma^2 \in \{0.005, 0.05\}$ \\
        
        \textbf{E}stimand & Mean of posterior dose-response function estimate $f(x)$ \\
        
        \textbf{M}ethods & \textbf{Non-linear functional shrinkage} (NLFS)  ($\theta_1 \sim N(0, 1), \, \sigma^2 \sim \text{IG}(0.001, 0.001)$)\\
        & - assuming Hill (NLFS (Hill)) \\
         & \quad Priors: $\theta_3 \sim N_+(0.5, 0.05)$, $\theta_4 \sim LN$ s.t. $\mathbb{E}(\theta_4) = 3$, $\mathbb{V}ar(\theta_4) = 3$ \\
        & - assuming power (NLFS (power)) \\
         & \quad Priors: $\theta_3 \sim N(0.5, 0.25)$ \\
        & - assuming Hill and power (NLFS (Hill+power)) \\
        & \quad Priors: As in NLFS(Hill) and NLFS(power) \\

        & \textbf{Parametric Bayesian fit} (Param.) ($\theta_1 \sim N(0, 1), \, \log(\sigma^2) \sim N(-1.75, 1) $)\\
        & - assuming Hill (Param.(Hill)) \\
        & \quad Priors: $\theta_3 \sim N_+(0.5, 0.05)$, $\theta_4 \sim LN$ s.t. $\mathbb{E}(\theta_4) = 3$, $\mathbb{V}ar(\theta_4) = 3$ \\
        & - assuming power (Param.(power)) \\
         & \quad Priors: $\theta \sim N(0.5, 0.25)$ \\
        
        & \textbf{B-spline} \\
        & \quad Priors: $\theta_1 \sim N(0, 1), \, \sigma^2 \sim \text{IG}(0.001, 0.001), \, \lambda^2 \sim \text{IG}(0.001, 0.001)$ \\
        & \textbf{P-spline} \\
        & \quad Priors: $\theta_1 \sim N(0, 1), \, \sigma^2 \sim \text{IG}(0.001, 0.001), \, \tau^2 \sim \text{IG}(1, 0.005)$ \\
        
        & \textbf{Parametric + horseshoe B-spline} \\
        & \quad $y = h_{\theta}(x) + \Phi(\beta) + \varepsilon$ \\
        & \quad Priors: \\
        & \quad $\beta \sim N(0, \sigma^2 \tau^2 \text{diag}(\lambda_1^2, \dots, \lambda_k^2))$ \\
        & \quad $\tau \sim C^{+}(0,1)$, $\lambda_j \overset{\text{iid}}{\sim} C^{+}(0, 1)$ \\
        & \quad $\theta_1 \sim N(0, 1)$, $\theta_2 \sim N(1.5, 2)$ (Scaling) \\
        & \quad - assuming Hill (Param.(Hill) + B-spline)) \\
        & \quad \quad Prior: $\theta_3 \sim N_+(0.5, 0.05)$, $\theta_4 \sim LN$ s.t. $\mathbb{E}(\theta_4) = 3$, $\mathbb{V}ar(\theta_4) = 3$ \\
        & \quad - assuming power (Param.(power) + B-spline) \\
        & \quad \quad Prior: $\theta \sim N(0.5, 0.25)$ \\
        \\
        \textbf{P}erformance & RMSE between posterior mean $\mathbb{E}(f(x)|d_s^i)$ and true $g(x)$ evaluated at drawn doses $x \in [0, 1]^n$
    \end{tabular}
\end{table}

\begin{sidewaystable}
    \centering
    \footnotesize
\caption{Simulation results for $\sigma^2=0.005$ ($2\sigma = 14.1\%$ of maximal effect) summarized by mean RMSE and corresponding standard deviation in parenthesis. OS means own slice and refers to setting $a$ and $b$ for the shrinkage parameters as suggested in \cite{shin2020functional} whereas HC means Half Cauchy and refers to the standard horseshoe prior with $a=b=0.5$.}
    \label{tab:sim_res_sigma_sq_0.0005}
\centering
\begin{tabular}{rlp{0.8cm}p{1cm}p{1cm}p{1cm}p{1cm}p{1cm}p{1cm}p{1cm}p{1cm}p{1cm}p{1cm}p{1cm}}
  \hline
  & & \multicolumn{3}{c}{n=50} & \multicolumn{3}{c}{n=100} & \multicolumn{3}{c}{n=200} & \multicolumn{3}{c}{n=500}\\
  \hline
& Method & Hill & power & Hill+down & Hill & power & Hill+down & Hill & power & Hill+down & Hill & power & Hill+down \\ 
  \hline
1 & NLFS(Hill), OS & 0.019 (0.007) & 0.027 (0.008) & 0.037 (0.008) & 0.014 (0.005) & 0.017 (0.005) & 0.028 (0.005) & 0.009 (0.003) & 0.011 (0.003) & 0.024 (0.002) & 0.006 (0.002) & 0.008 (0.002) & 0.017 (0.004) \\ 
  2 & NLFS(power), OS & 0.046 (0.013) & 0.018 (0.006) & 0.053 (0.017) & 0.031 (0.005) & 0.013 (0.005) & 0.031 (0.005) & 0.023 (0.004) & 0.009 (0.003) & 0.022 (0.004) & 0.015 (0.002) & 0.006 (0.002) & 0.015 (0.002) \\ 
  3 & NLFS(Hill+power), OS & 0.021 (0.007) & 0.022 (0.007) & 0.028 (0.006) & 0.015 (0.005) & 0.015 (0.005) & 0.023 (0.004) & 0.01 (0.003) & 0.011 (0.003) & 0.018 (0.003) & 0.007 (0.002) & 0.007 (0.002) & 0.011 (0.003) \\ 
  4 & NLFS(Hill), HC & 0.019 (0.007) & 0.022 (0.007) & 0.033 (0.007) & 0.014 (0.005) & 0.015 (0.005) & 0.027 (0.004) & 0.01 (0.003) & 0.011 (0.003) & 0.024 (0.002) & 0.006 (0.002) & 0.008 (0.002) & 0.015 (0.004) \\ 
  5 & NLFS(power), HC & 0.046 (0.015) & 0.017 (0.006) & 0.054 (0.019) & 0.03 (0.005) & 0.012 (0.005) & 0.03 (0.005) & 0.021 (0.004) & 0.009 (0.003) & 0.022 (0.004) & 0.013 (0.002) & 0.006 (0.002) & 0.013 (0.002) \\ 
  6 & NLFS(Hill\&power), HC & 0.021 (0.007) & 0.021 (0.007) & 0.028 (0.006) & 0.015 (0.005) & 0.015 (0.005) & 0.023 (0.004) & 0.01 (0.003) & 0.011 (0.003) & 0.018 (0.003) & 0.007 (0.002) & 0.007 (0.002) & 0.011 (0.002) \\ 
  7 & param(Hill) & 0.019 (0.007) & 0.023 (0.005) & 0.052 (0.008) & 0.013 (0.005) & 0.019 (0.004) & 0.05 (0.005) & 0.009 (0.003) & 0.016 (0.002) & 0.05 (0.003) & 0.006 (0.002) & 0.012 (0.001) & 0.049 (0.002) \\ 
  8 & param(power) & 0.151 (0.011) & 0.015 (0.006) & 0.18 (0.011) & 0.152 (0.008) & 0.011 (0.005) & 0.18 (0.008) & 0.154 (0.005) & 0.008 (0.003) & 0.181 (0.006) & 0.154 (0.003) & 0.005 (0.002) & 0.182 (0.004) \\ 
  9 & bspline & 0.042 (0.007) & 0.042 (0.008) & 0.042 (0.007) & 0.032 (0.005) & 0.033 (0.005) & 0.032 (0.005) & 0.023 (0.004) & 0.024 (0.004) & 0.023 (0.004) & 0.014 (0.002) & 0.014 (0.002) & 0.014 (0.002) \\ 
  10 & pspline & 0.03 (0.007) & 0.026 (0.006) & 0.03 (0.007) & 0.022 (0.005) & 0.019 (0.004) & 0.022 (0.005) & 0.016 (0.003) & 0.015 (0.003) & 0.016 (0.003) & 0.011 (0.002) & 0.01 (0.002) & 0.011 (0.002) \\ 
  11 & param(Hill)+bspline & 0.021 (0.007) & 0.022 (0.006) & 0.031 (0.006) & 0.015 (0.005) & 0.017 (0.004) & 0.024 (0.005) & 0.011 (0.003) & 0.014 (0.003) & 0.018 (0.003) & 0.007 (0.002) & 0.01 (0.002) & 0.012 (0.002) \\ 
  12 & param(power)+bspline & 0.038 (0.007) & 0.019 (0.007) & 0.04 (0.007) & 0.029 (0.005) & 0.013 (0.005) & 0.03 (0.005) & 0.021 (0.004) & 0.009 (0.003) & 0.022 (0.004) & 0.013 (0.002) & 0.006 (0.002) & 0.014 (0.002) \\
     \hline
\end{tabular}
\end{sidewaystable}

\begin{sidewaystable}
    \centering
    \footnotesize
\caption{Simulation results for $\sigma^2=0.05$ ($2\sigma = 44.7\%$ of maximal effect) summarized by mean RMSE and corresponding standard deviation in parenthesis. OS means own slice and refers to setting $a$ and $b$ for the shrinkage parameters as suggested in \cite{shin2020functional} whereas HC means Half Cauchy and refers to the standard horseshoe prior with $a=b=0.5$.}
    \label{tab:sim_res_sigma_sq_0.005}
\centering
\begin{tabular}{rlp{0.8cm}p{1cm}p{1cm}p{1cm}p{1cm}p{1cm}p{1cm}p{1cm}p{1cm}p{1cm}p{1cm}p{1cm}}
  \hline
  & & \multicolumn{3}{c}{n=50} & \multicolumn{3}{c}{n=100} & \multicolumn{3}{c}{n=200} & \multicolumn{3}{c}{n=500}\\
  \hline
& Method & Hill & power & Hill+down & Hill & power & Hill+down & Hill & power & Hill+down & Hill & power & Hill+down \\ 
  \hline
1 & NLFS(Hill), OS & 0.059 (0.021) & 0.081 (0.021) & 0.076 (0.017) & 0.043 (0.015) & 0.058 (0.015) & 0.063 (0.012) & 0.03 (0.01) & 0.039 (0.011) & 0.051 (0.009) & 0.019 (0.007) & 0.022 (0.007) & 0.036 (0.007) \\ 
  2 & NLFS(power), OS & 0.137 (0.015) & 0.053 (0.019) & 0.155 (0.022) & 0.107 (0.023) & 0.039 (0.014) & 0.101 (0.024) & 0.066 (0.012) & 0.027 (0.011) & 0.067 (0.012) & 0.042 (0.007) & 0.017 (0.007) & 0.043 (0.007) \\ 
  3 & NLFS(Hill+power), OS & 0.064 (0.021) & 0.066 (0.021) & 0.072 (0.019) & 0.046 (0.015) & 0.046 (0.015) & 0.055 (0.014) & 0.032 (0.01) & 0.031 (0.011) & 0.041 (0.01) & 0.02 (0.007) & 0.02 (0.007) & 0.03 (0.006) \\ 
  4 & NLFS(Hill), HC & 0.057 (0.02) & 0.067 (0.02) & 0.073 (0.017) & 0.042 (0.015) & 0.049 (0.015) & 0.059 (0.013) & 0.03 (0.01) & 0.034 (0.011) & 0.046 (0.01) & 0.02 (0.007) & 0.021 (0.007) & 0.033 (0.006) \\ 
  5 & NLFS(power), HC & 0.133 (0.016) & 0.048 (0.02) & 0.141 (0.026) & 0.098 (0.021) & 0.036 (0.015) & 0.097 (0.019) & 0.066 (0.013) & 0.026 (0.011) & 0.072 (0.014) & 0.042 (0.007) & 0.017 (0.007) & 0.043 (0.008) \\ 
  6 & NLFS(Hill+power), HC & 0.061 (0.02) & 0.06 (0.019) & 0.068 (0.019) & 0.045 (0.015) & 0.045 (0.015) & 0.053 (0.014) & 0.032 (0.01) & 0.032 (0.011) & 0.041 (0.01) & 0.02 (0.007) & 0.02 (0.007) & 0.03 (0.006) \\ 
  7 & param(Hill) & 0.063 (0.021) & 0.054 (0.018) & 0.082 (0.016) & 0.043 (0.016) & 0.042 (0.013) & 0.066 (0.011) & 0.03 (0.011) & 0.033 (0.009) & 0.058 (0.006) & 0.019 (0.007) & 0.024 (0.005) & 0.053 (0.003) \\ 
  8 & param(power) & 0.161 (0.012) & 0.043 (0.019) & 0.188 (0.012) & 0.157 (0.008) & 0.032 (0.014) & 0.184 (0.009) & 0.156 (0.005) & 0.024 (0.01) & 0.183 (0.006) & 0.155 (0.003) & 0.016 (0.007) & 0.183 (0.004) \\ 
  9 & bspline & 0.112 (0.02) & 0.107 (0.022) & 0.111 (0.02) & 0.088 (0.014) & 0.084 (0.014) & 0.086 (0.014) & 0.068 (0.01) & 0.067 (0.01) & 0.067 (0.01) & 0.047 (0.006) & 0.048 (0.006) & 0.046 (0.006) \\ 
  10 & pspline & 0.081 (0.019) & 0.063 (0.017) & 0.08 (0.019) & 0.061 (0.014) & 0.048 (0.013) & 0.06 (0.013) & 0.045 (0.01) & 0.037 (0.009) & 0.045 (0.01) & 0.03 (0.006) & 0.026 (0.006) & 0.03 (0.006) \\ 
  11 & param(Hill)+bspline, HC & 0.072 (0.021) & 0.06 (0.02) & 0.081 (0.019) & 0.05 (0.016) & 0.045 (0.014) & 0.061 (0.014) & 0.034 (0.011) & 0.034 (0.01) & 0.047 (0.009) & 0.022 (0.007) & 0.023 (0.006) & 0.033 (0.006) \\ 
  12 & param(power)+bspline, HC & 0.097 (0.019) & 0.053 (0.02) & 0.103 (0.019) & 0.076 (0.014) & 0.039 (0.015) & 0.081 (0.014) & 0.058 (0.01) & 0.028 (0.01) & 0.061 (0.01) & 0.04 (0.006) & 0.019 (0.007) & 0.042 (0.006) \\ 
     \hline
\end{tabular}
\end{sidewaystable}

\section{Figures}

\begin{figure}
    \centering
    \includegraphics[width = 1\linewidth]{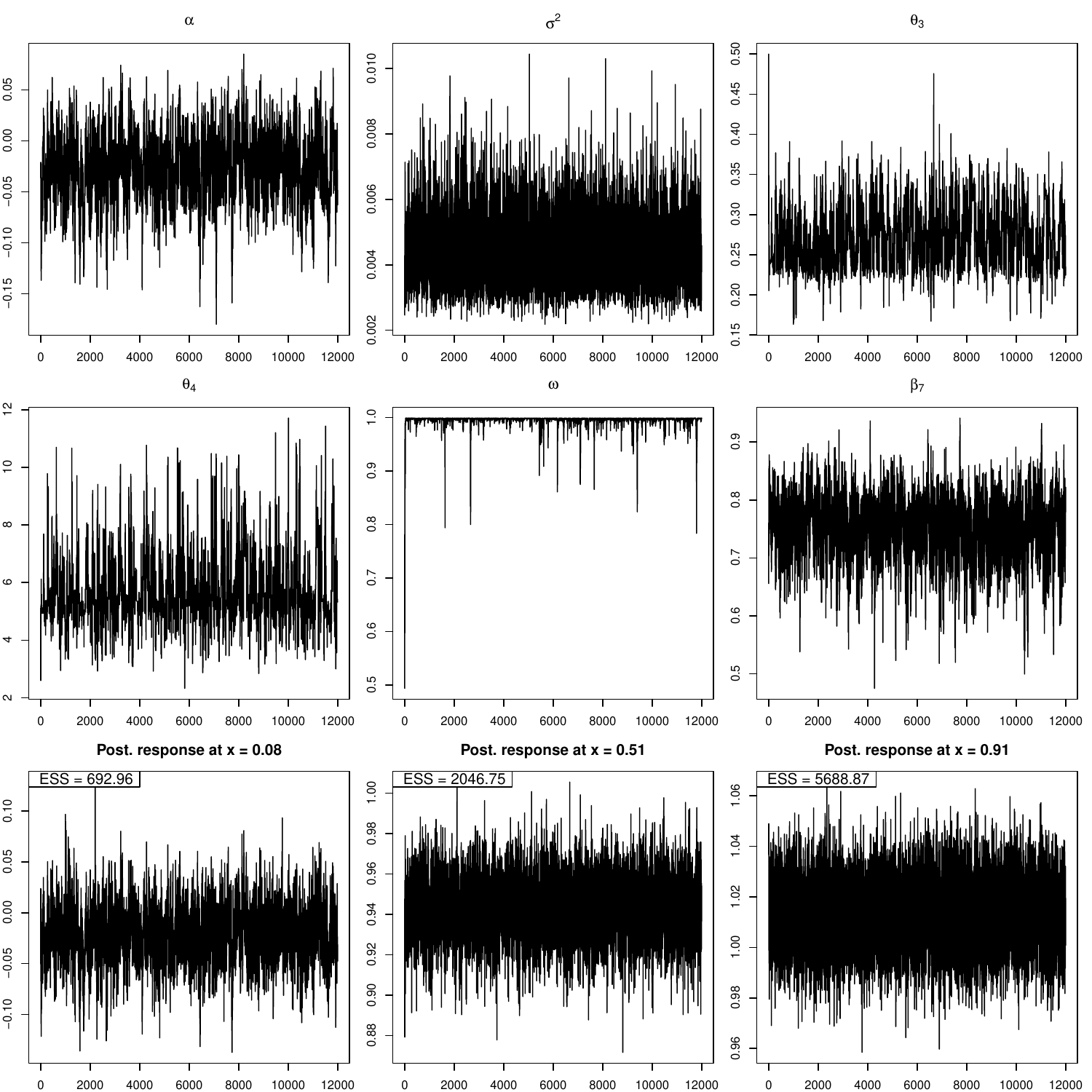}
    \caption{Traceplots of the NLFS(Hill) example fit in Figure \ref{fig:examples}, pane (a), correct subspace specification. The first 2000 samples were discarded as burn-in.
    Due to the correct subspace specification, there is strong shrinkage ($\omega = (1 + \tau^2)^{-1}$ close to 1). The effective sample size (ESS) was calculated based on the 10000 draws after discarding the first 2000 burn-in draws using the \texttt{coda} R-package \citep{coda}.}
    \label{fig:traceplots_correct}
\end{figure}

\begin{figure}
    \centering
    \includegraphics[width = 1\linewidth]{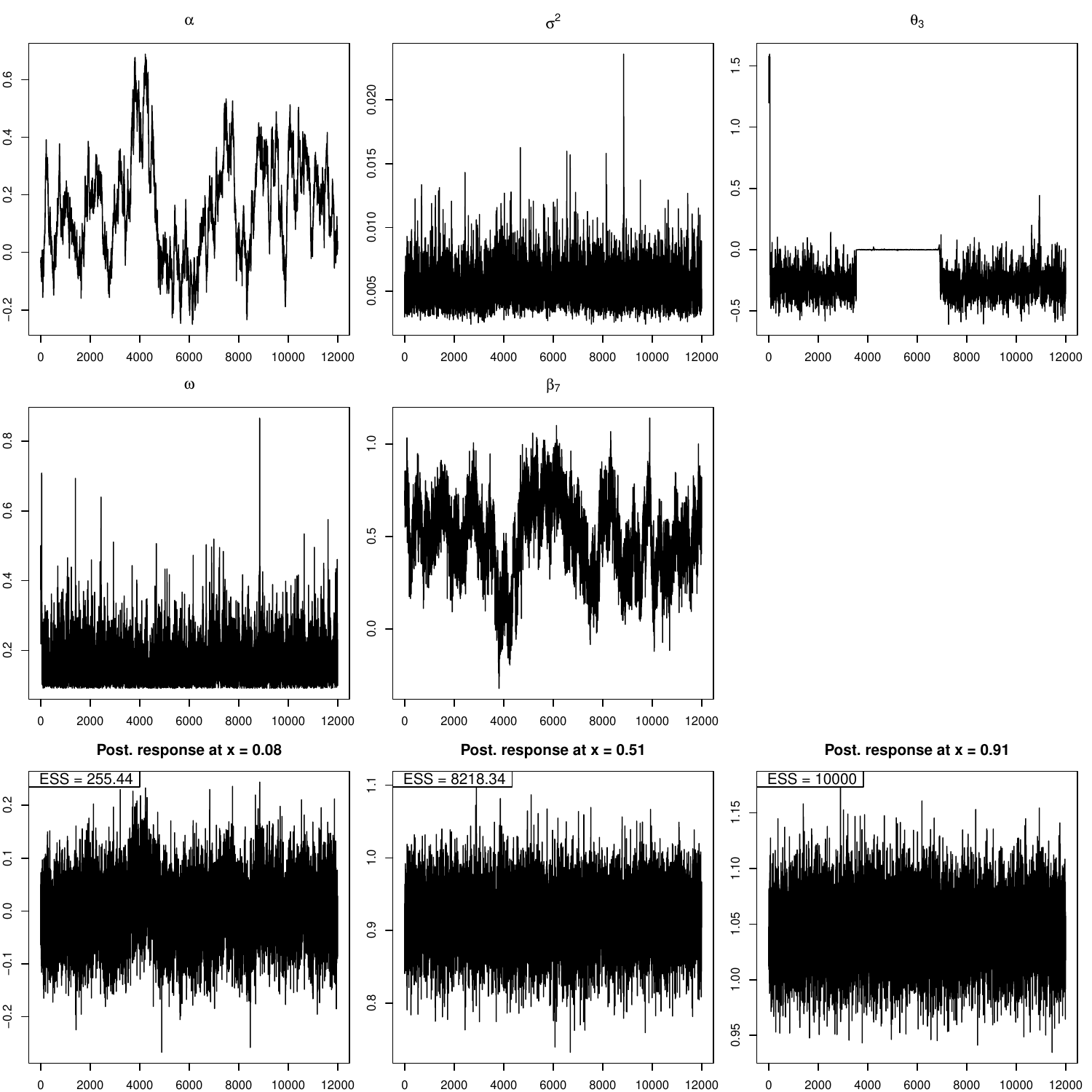}
    \caption{Traceplots of the NLFS(power) example fit in Figure \ref{fig:examples}, pane (b), subspace misspecification. The first 2000 samples were discarded as burn-in. Due to the subspace misspecification, there is little shrinkage ($\omega = (1 + \tau^2)^{-1}$ close to 0). Further, power exponent $\theta_3$ is stuck at 0 for a few thousand draws, and $\theta_1$, the intercept, seems highly correlated. Due to the misspecification and effectively no shrinkage, single parameters are not well identifiable and the whole response must be viewed to inspect convergence. The resulting response mixes well (bottom row). The effective sample size (ESS) was calculated based on the 10000 draws after discarding the first 2000 burn-in draws using the \texttt{coda} R-package \citep{coda}.}
    \label{fig:traceplots_misspec}
\end{figure}

\section{Proofs} 

\textbf{Lemma 1:} $P_{\theta}$ does not depend on linear parameters.

\begin{proof}
Let $h(x, \theta) = \theta_1 + \theta_2 q(x, \theta_3)$ be a twice differentiable function  with $q$ non-linear in $\theta_3$. W.L.o.G. assume that $\theta_3 \in \mathbb{R}$. Then
\[\dot{\text{H}}_{\theta} = \frac{\partial h(x, \theta)}{\partial \theta} = (1_n \quad \underbrace{q(x, \theta_3)}_{:=c_1} \quad \theta_2 \underbrace{\frac{\partial q(x, \theta_3)}{\partial \theta_3}}_{:=c_2}) = \underbrace{(1_n \quad c_1 \quad c_2)}_{:=H_1 \in \mathbb{R}^{n\times 3}} \underbrace{\begin{pmatrix}
    1 & 0 & 0 \\
    0 & 1 & 0 \\
    0 & 0 & \theta_2
\end{pmatrix}}_{H_2 \in \mathbb{R}^{3\times3}} \]
and 
\begin{align*}
    P_{\theta} &= \dot{\text{H}}_{\theta} (\dot{\text{H}}_{\theta}^{\top} \dot{\text{H}}_{\theta})^{-1}\dot{\text{H}}_{\theta}^{\top} \\
    &= H_1H_2(H_2 H_1^{\top}H_1H_2)^{-1}H_2H_1^{\top}\\
    &= H_1 H_2 H_2^{-1} (H_1^{\top}H_1)^{-1}H_2^{-1}H_2H_1^{\top} \\
    &= H_1 (H_1^{\top}H_1)^{-1} H_1^{\top}
\end{align*}

and $H_1$ does not depend on $\theta_2$. \qedsymbol
\end{proof}

\section{Computation} 

The code to reproduce results is available at \newline \href{https://gitlab.tu-dortmund.de/functional_shrinkage/nonlinear_shrinkage}{https://gitlab.tu-dortmund.de/functional\_shrinkage/nonlinear\_shrinkage}.

The non-linear functional shrinkage (NLFS) approach for the Hill model is implemented using a combination of Gibbs-, Metropolis-Hastings- and Slice sampling \citep{brooks2011handbook, neal2003slice}.
We separately model the function intercept $\theta_1$. 

Given the likelihood and priors
		\begin{align*}
			Y &\sim N(\theta_1 1_n +  \Phi \beta, \sigma^2 I_n)\\ 
			\sigma^2 & \sim \text{IG}(a_\sigma, b_\sigma), \, 
			\theta_1 \sim N(\mu_{\theta_1},\sigma^2_{\theta_1}) \\
			\beta & \sim N(0, \sigma^2 \tau^2 (\Phi^{\top}(I - P_{\theta})\Phi)^{-1}) \\
			\theta_3 & \sim N_{+}(\mu_{\theta_3}, \sigma^2_{\theta_3}), \, 
			\theta_4 \sim LN(\mu_{\theta_4}, \sigma^2_{\theta_4}) \\
			\omega &= 1/(1+\tau^2)  \sim \text{Beta}(a_{\omega}, b_{\omega}), \,
			a_{\omega}=0.5, \, b_{\omega}=\exp(-\log(n)/2), \\
		\end{align*}

the sampler is summarized in Algorithm \ref{algorithm_sampler}.

\begin{algorithm}
	\caption{Non-linear functional shrinkage (NLFS)}\label{algorithm_sampler}
	\begin{algorithmic}[1]
		\Initialize{$\beta^{(1)}, \sigma^{2(1)}, \tau^{(1)}, \omega^{(1)}, \theta^{(1)} $}
		\For{$i: \, 2 \rightarrow B$}
		\State	\textbf{Calculate} $\dot{\text{H}}_{\theta^{(i-1)}}$
		\State \textbf{Sample} $\beta^{(i)} \sim p(\beta|\cdot)$ \Comment{Conjugate}
		\State \textbf{Sample} $\theta_1^{(i)} \sim p(\theta_1|\cdot)$ \Comment{Conjugate}
		\State \textbf{Sample} $\sigma^{2(i)} \sim p(\sigma^2|\cdot)$ \Comment{Conjugate}
		\State \textbf{Sample} $\omega^{(i)} \sim p(\omega|\cdot)$ \Comment{Slice Sampler}
		\State \textbf{Sample} Non-linear $\theta^{(i)} \sim p(\theta|\cdot)$ \Comment{MH Sampler}
		\EndFor
		\State \textbf{return} All samples
	\end{algorithmic}
\end{algorithm}

\textbf{Update $\beta$} \\
We update $\beta$ using the full conditional posterior
\begin{equation}
    \beta | \sigma^2, \tau^2, \theta, y \sim N(\mu_{\beta}', \Sigma_{\beta}')
\end{equation}
where $\Sigma_{\beta}' = \sigma^2(\Phi^{\top} \Phi + \tau^2 \Phi^{\top}(I_n - P_{\theta})\Phi)^{-1}$ and $\mu_{\beta}' = \sigma^{-2} \Sigma_{\beta}' \Phi' \tilde y$ and $\tilde y = y - 1_n\theta_1$. 

\textbf{Update $\theta_1$} \\

The intercept $\theta_1$ is updated using
\begin{equation}
    \theta_1| \beta, \sigma^2, y \sim N(\mu_{\theta_1}', {\sigma^2_{\theta_1}}')
\end{equation}
where ${\sigma^2_{\theta_1}}' = (\sigma^2 \sigma^2_{\theta_1}) (n \sigma^2_{\theta_1} + \sigma^2)$ and $\mu_{\theta_1}' = \sigma^{-2} {\sigma^2_{\theta_1}}' 1_n^{\top}(y-\Phi \beta) + \mu_{\theta_1} / \sigma^2_{\theta_1}$. 

\textbf{Update $\sigma^2$} \\
The noise variance $\sigma^2$ is updated by
\begin{equation}
    \sigma^2 | \beta, \tau^2, \theta \sim \text{IG}(a_\sigma', b_\sigma') 
\end{equation}%
where $a_{\sigma}' = (n+k)/2 + a_{\sigma}$ and $b_{\sigma}' = 0.5 (\text{RSS} + \tau^{-2} \beta^{\top} (\Phi^{\top}(I_n - P_{\theta})\Phi) \beta) + b_{\sigma}$ where $RSS= {||y - (\theta_1 1_n  + \Phi \beta) ||}_2^2$ is the residual sum of squares and $||.||_2$ is the Euclidean norm.

\textbf{Update $\tau^2$} \\
%
We update $\tau^2$ using a slice sampler \citep{neal2003slice} considering the posterior log likelihood
\begin{align}
g(\tau) = \log(p(\tau|\beta, \sigma^2, \theta)) =& (-k/2 + b_{\omega} - 0.5) \log(\tau^2) \\
& + (-a_{\omega} - b_{\omega}) \log(1+\tau^2) \\
& + \left(-\frac{1}{2\sigma^2} \beta^{\top}\Phi^{\top}(I_n - P_{\theta})\Phi \beta\right) \tau^{-2}.
\end{align}%
Note that we use $-k/2$ and not $-(k-d_0)/2$ as in \cite{shin2020functional} where $d_0$ is the rank of the (linear) projection matrix.
We omit $-d_0$ as for the non-linear approach, there are no linear bases in $\dot{\text{H}}_{\theta}$ that are in $\Phi$ (because there is no intercept in $\Phi$) and hence the prior covariance matrix of $\beta$ is of full rank $k$.
For a current $\tau_{0}$, calculate $v=g(\tau_0))$.
Uniformly draw $\tilde z \sim U(0, \exp(v))$. For $z=\log(\tilde{z})$ define the slice $S_z = \{x: g(x) < g(z)\}$ and sample the next $\tau_1$ uniformly from $S_z$. For computational ease, we restrict $\tau^2$ to [0.001, 10].

The above sampling for the general $\omega \sim \text{Beta}(a_{\omega}, b_{\omega})$ prior was primarily featured and labelled 'own slice' (OC) in Table \ref{tab:sim_res_sigma_sq_0.0005}. We also considered a standard horseshoe (HS) prior ($a=b=0.5$).
Details on its implementation are in \cite{makalic2015simple}.

\textbf{Update non-linear $\theta$} \\
One only has to update the non-linear parameters of $\theta$, as $P_{\theta}$ only depends on the non-linear parameters.
For the Hill model, the non-linear parameters are $\theta_3$ and $\theta_4$.
We assume independence and separately update $\theta_3$ and $\theta_4$ using a Metropolis-Hastings sampler \citep{brooks2011handbook} and explain the sampling for $\theta_3$.

Perform the three sampling steps
\begin{enumerate}
    \item Draw a candidate $\theta_3^{(1)}$ from a proposal distribution $p_{\text{prop}} $using $\theta_3^{(0)}$
    \item Calculate the hastings ratio
    \[\text{HR} = \frac{p(\theta_{3}^{(1)} | \cdot) p_{\text{prop}}(\theta_3^{(0)} |\theta_3^{(1)})}{p(\theta_3^{(0)} | \cdot) p_{\text{prop}}(\theta_3^{(1)} | \theta_3^{(0)})}.\]
    \item Draw $u \sim \text{Unif}[0,1]$. If $\text{HR} > u$, accept $\theta_3^{(1)}$ as new draw. Otherwise, reject and consider $\theta_3^{(0)}$ as new draw.
    
\end{enumerate}

For step (1),  sample a new candidate $\theta_3^{(1)}$ from a proposal distribution, e.g. $N_{+}(\theta_3^{(0)}, \sigma^2_{\text{prop}})$ where $\sigma^2_{\text{prop}}$ might be calculated as, e.g. the empirical variance of the latest 100 draws of $\theta_3$, or simply as  $\sigma^2_{\text{prop}} = \sigma^2_{\theta_3}$.
To sample $x$ from a truncated normal distribution with positive support, $x \sim N_{+}(\mu, \sigma^2)$, calculate $l = P(X < 0)$ where $X \sim N(\mu, \sigma^2)$. Sample $u \sim \text{Unif}[l, 1]$ and calculate $x = q_{\mu, \sigma^2}(u)$, the corresponding quantile.

For step (2), consider $\log(\text{HR})$ for computational stability:
\[\log(\text{HR}) = \log(p(\theta_3^{(1)} | \cdot)) - \log(p(\theta_3^{(0)} | \cdot)) + \log(p_{\text{prop}}(\theta_3^{(0)} | \theta_3^{(1)})) - \log(p_{\text{prop}}(\theta_3^{(1)} | \theta_3^{(0)})). \]
For the fully conditional log posterior $\log(p(\theta | \cdot))$, we can integrate out $\beta$ to reduce the autocorrelation in the sampling. Since $p(\theta | \cdot) \propto p(y | \theta, \sigma^2, \tau^2) p(\theta)$ and \[y | \theta, \sigma^2, \tau^2 \sim N(\mathbb{E}(\Phi\beta + \varepsilon), \text{Cov}(\Phi\beta + \varepsilon)),\]

we use $y|\theta, \sigma^2, \tau^2 \sim N(0, \Sigma_{y})$ with $\Sigma_y = \sigma^2 \tau^2 \Phi (\Phi^{\top}(I-P_{\theta})\Phi)^{-1}\Phi^{\top} + \sigma^2I_n$.

\newpage


\begin{thebibliography}{33}
	\providecommand{\natexlab}[1]{#1}
	\providecommand{\url}[1]{\texttt{#1}}
	\expandafter\ifx\csname urlstyle\endcsname\relax
	\providecommand{\doi}[1]{doi: #1}\else
	\providecommand{\doi}{doi: \begingroup \urlstyle{rm}\Url}\fi
	
	\bibitem[Alvarez et~al.(2013)Alvarez, Luengo, and Lawrence]{alvarez2013linear}
	M.~A. Alvarez, D.~Luengo, and N.~D. Lawrence.
	\newblock Linear latent force models using gaussian processes.
	\newblock \emph{IEEE transactions on pattern analysis and machine
		intelligence}, 35\penalty0 (11):\penalty0 2693--2705, 2013.
	
	\bibitem[Brezger and Steiner(2008)]{brezger2008}
	A.~Brezger and W.~J. Steiner.
	\newblock Monotonic regression based on bayesian p–splines.
	\newblock \emph{Journal of Business \& Economic Statistics}, 26\penalty0
	(1):\penalty0 90--104, 2008.
	\newblock \doi{10.1198/073500107000000223}.
	\newblock URL \url{https://doi.org/10.1198/073500107000000223}.
	
	\bibitem[Brooks et~al.(2011)Brooks, Gelman, Jones, and
	Meng]{brooks2011handbook}
	S.~Brooks, A.~Gelman, G.~Jones, and X.-L. Meng.
	\newblock \emph{Handbook of markov chain monte carlo}.
	\newblock CRC press, 2011.
	
	\bibitem[Carl(2001)]{carl2001practical}
	D.~Carl.
	\newblock \emph{A practical guide to splines}.
	\newblock Springer, 2001.
	
	\bibitem[Carvalho et~al.(2010)Carvalho, Polson, and
	Scott]{carvalho2010horseshoe}
	C.~M. Carvalho, N.~G. Polson, and J.~G. Scott.
	\newblock The horseshoe estimator for sparse signals.
	\newblock \emph{Biometrika}, 97\penalty0 (2):\penalty0 465--480, 2010.
	
	\bibitem[Chen et~al.(2022)Chen, Chen, Zhang, and Jeff~Wu]{chen2022apik}
	J.~Chen, Z.~Chen, C.~Zhang, and C.~Jeff~Wu.
	\newblock Apik: Active physics-informed kriging model with partial differential
	equations.
	\newblock \emph{SIAM/ASA Journal on Uncertainty Quantification}, 10\penalty0
	(1):\penalty0 481--506, 2022.
	
	\bibitem[Golightly and Wilkinson(2011)]{golightly2011bayesian}
	A.~Golightly and D.~J. Wilkinson.
	\newblock Bayesian parameter inference for stochastic biochemical network
	models using particle markov chain monte carlo.
	\newblock \emph{Interface focus}, 1\penalty0 (6):\penalty0 807--820, 2011.
	
	\bibitem[Goutelle et~al.(2008)Goutelle, Maurin, Rougier, Barbaut, Bourguignon,
	Ducher, and Maire]{goutelle2008hill}
	S.~Goutelle, M.~Maurin, F.~Rougier, X.~Barbaut, L.~Bourguignon, M.~Ducher, and
	P.~Maire.
	\newblock The hill equation: a review of its capabilities in pharmacological
	modelling.
	\newblock \emph{Fundamental \& clinical pharmacology}, 22\penalty0
	(6):\penalty0 633--648, 2008.
	
	\bibitem[Gunn and Dunson(2005)]{gunn2005transformation}
	L.~H. Gunn and D.~B. Dunson.
	\newblock A transformation approach for incorporating monotone or unimodal
	constraints.
	\newblock \emph{Biostatistics}, 6\penalty0 (3):\penalty0 434--449, 2005.
	
	\bibitem[Hill(1910)]{hill1910possible}
	A.~V. Hill.
	\newblock The possible effects of the aggregation of the molecules of
	hemoglobin on its dissociation curves.
	\newblock \emph{The Journal of Physiology}, 40:\penalty0 iv--vii, 1910.
	
	\bibitem[Hoerl and Kennard(1970)]{hoerl1970ridge}
	A.~E. Hoerl and R.~W. Kennard.
	\newblock Ridge regression: Biased estimation for nonorthogonal problems.
	\newblock \emph{Technometrics}, 12\penalty0 (1):\penalty0 55--67, 1970.
	
	\bibitem[Huang et~al.(2006)Huang, Liu, and Wu]{huang2006hierarchical}
	Y.~Huang, D.~Liu, and H.~Wu.
	\newblock Hierarchical bayesian methods for estimation of parameters in a
	longitudinal hiv dynamic system.
	\newblock \emph{Biometrics}, 62\penalty0 (2):\penalty0 413--423, 2006.
	
	\bibitem[Kelsey et~al.(2014)Kelsey, Li, Mitchell, Whelan, Anderson, and
	Wallace]{kelsey2014validated}
	T.~W. Kelsey, L.~Q. Li, R.~T. Mitchell, A.~Whelan, R.~A. Anderson, and W.~H.~B.
	Wallace.
	\newblock A validated age-related normative model for male total testosterone
	shows increasing variance but no decline after age 40 years.
	\newblock \emph{PloS one}, 9\penalty0 (10):\penalty0 e109346, 2014.
	
	\bibitem[K{\"o}llmann et~al.(2014)K{\"o}llmann, Bornkamp, and
	Ickstadt]{kollmann2014unimodal}
	C.~K{\"o}llmann, B.~Bornkamp, and K.~Ickstadt.
	\newblock Unimodal regression using bernstein--schoenberg splines and
	penalties.
	\newblock \emph{Biometrics}, 70\penalty0 (4):\penalty0 783--793, 2014.
	
	\bibitem[Lang and Brezger(2004)]{lang2004}
	S.~Lang and A.~Brezger.
	\newblock Bayesian p-splines.
	\newblock \emph{Journal of computational and graphical statistics}, 13\penalty0
	(1):\penalty0 183--212, 2004.
	
	\bibitem[Lunn et~al.(2002)Lunn, Best, Thomas, Wakefield, and
	Spiegelhalter]{lunn2002bayesian}
	D.~J. Lunn, N.~Best, A.~Thomas, J.~Wakefield, and D.~Spiegelhalter.
	\newblock Bayesian analysis of population pk/pd models: general concepts and
	software.
	\newblock \emph{Journal of pharmacokinetics and pharmacodynamics}, 29:\penalty0
	271--307, 2002.
	
	\bibitem[Makalic and Schmidt(2015)]{makalic2015simple}
	E.~Makalic and D.~F. Schmidt.
	\newblock A simple sampler for the horseshoe estimator.
	\newblock \emph{IEEE Signal Processing Letters}, 23\penalty0 (1):\penalty0
	179--182, 2015.
	
	\bibitem[Mesarovic et~al.(2004)Mesarovic, Sreenath, and
	Keene]{mesarovic2004search}
	D.~Mesarovic, Mihajlo, S.~Sreenath, and J.~Keene.
	\newblock Search for organising principles: understanding in systems biology.
	\newblock \emph{Systems biology}, 1\penalty0 (1):\penalty0 19--27, 2004.
	
	\bibitem[Meyer(2008)]{mayer2008}
	M.~C. Meyer.
	\newblock Inference using shape-restricted regression splines.
	\newblock \emph{The Annals of Applied Statistics}, 2\penalty0 (3):\penalty0
	1013--1033, 2008.
	\newblock ISSN 19326157.
	\newblock URL \url{http://www.jstor.org/stable/30245118}.
	
	\bibitem[Meyer et~al.(2011)Meyer, Hackstadt, and Hoeting]{meyer2011}
	M.~C. Meyer, A.~J. Hackstadt, and J.~A. Hoeting.
	\newblock Bayesian estimation and inference for generalised partial linear
	models using shape-restricted splines.
	\newblock \emph{Journal of Nonparametric Statistics}, 23\penalty0 (4):\penalty0
	867--884, 2011.
	
	\bibitem[Morris et~al.(2019)Morris, White, and Crowther]{morris2019using}
	T.~P. Morris, I.~R. White, and M.~J. Crowther.
	\newblock Using simulation studies to evaluate statistical methods.
	\newblock \emph{Statistics in medicine}, 38\penalty0 (11):\penalty0 2074--2102,
	2019.
	
	\bibitem[Neal(2003)]{neal2003slice}
	R.~M. Neal.
	\newblock Slice sampling.
	\newblock \emph{The annals of statistics}, 31\penalty0 (3):\penalty0 705--767,
	2003.
	
	\bibitem[Plummer et~al.(2006)Plummer, Best, Cowles, and Vines]{coda}
	M.~Plummer, N.~Best, K.~Cowles, and K.~Vines.
	\newblock Coda: Convergence diagnosis and output analysis for {MCMC}.
	\newblock \emph{R News}, 6\penalty0 (1):\penalty0 7--11, 2006.
	\newblock URL \url{https://journal.r-project.org/archive/}.
	
	\bibitem[Seber and Wild(2003)]{seber2003nonlinear}
	G.~A. Seber and C.~J. Wild.
	\newblock Nonlinear regression. hoboken.
	\newblock \emph{New Jersey: John Wiley \& Sons}, 62\penalty0 (63):\penalty0
	1238, 2003.
	
	\bibitem[Shin et~al.(2020)Shin, Bhattacharya, and Johnson]{shin2020functional}
	M.~Shin, A.~Bhattacharya, and V.~E. Johnson.
	\newblock Functional horseshoe priors for subspace shrinkage.
	\newblock \emph{Journal of the American Statistical Association}, 115\penalty0
	(532):\penalty0 1784--1797, 2020.
	
	\bibitem[Shively et~al.(2009)Shively, Sager, and Walker]{shivley2009}
	T.~S. Shively, T.~W. Sager, and S.~G. Walker.
	\newblock A bayesian approach to non-parametric monotone function estimation.
	\newblock \emph{Journal of the Royal Statistical Society. Series B (Statistical
		Methodology)}, 71\penalty0 (1):\penalty0 159--175, 2009.
	
	\bibitem[Shively et~al.(2011)Shively, Walker, and Damien]{shively2011}
	T.~S. Shively, S.~G. Walker, and P.~Damien.
	\newblock Nonparametric function estimation subject to monotonicity, convexity
	and other shape constraints.
	\newblock \emph{Journal of Econometrics}, 161\penalty0 (2):\penalty0 166--181,
	2011.
	
	\bibitem[{\v{S}}imon(2005)]{vsimon2005considerations}
	P.~{\v{S}}imon.
	\newblock Considerations on the single-step kinetics approximation.
	\newblock \emph{Journal of Thermal Analysis and Calorimetry}, 82\penalty0
	(3):\penalty0 651--657, 2005.
	
	\bibitem[Tibshirani(1996)]{tibshirani1996regression}
	R.~Tibshirani.
	\newblock Regression shrinkage and selection via the lasso.
	\newblock \emph{Journal of the Royal Statistical Society: Series B
		(Methodological)}, 58\penalty0 (1):\penalty0 267--288, 1996.
	
	\bibitem[Titsias et~al.(2012)Titsias, Honkela, Lawrence, and
	Rattray]{titsias2012identifying}
	M.~K. Titsias, A.~Honkela, N.~D. Lawrence, and M.~Rattray.
	\newblock Identifying targets of multiple co-regulating transcription factors
	from expression time-series by bayesian model comparison.
	\newblock \emph{BMC systems biology}, 6:\penalty0 1--21, 2012.
	
	\bibitem[Wheeler et~al.(2014)Wheeler, Dunson, Pandalai, Baker, and
	Herring]{wheeler2014mechanistic}
	M.~W. Wheeler, D.~B. Dunson, S.~P. Pandalai, B.~A. Baker, and A.~H. Herring.
	\newblock Mechanistic hierarchical gaussian processes.
	\newblock \emph{Journal of the American Statistical Association}, 109\penalty0
	(507):\penalty0 894--904, 2014.
	
	\bibitem[Wheeler et~al.(2017)Wheeler, Dunson, and Herring]{wheeler2017bayesian}
	M.~W. Wheeler, D.~B. Dunson, and A.~H. Herring.
	\newblock Bayesian local extremum splines.
	\newblock \emph{Biometrika}, 104\penalty0 (4):\penalty0 939--952, 2017.
	
	\bibitem[Wiemann and Kneib(2021)]{wiemann2021adaptive}
	P.~Wiemann and T.~Kneib.
	\newblock Adaptive shrinkage of smooth functional effects towards a predefined
	functional subspace.
	\newblock \emph{arXiv preprint arXiv:2101.05630}, 2021.
	
\end{thebibliography}

\end{document}